\documentclass{ws-procs975x65}

\def\beq{\begin{equation}}
\def\eeq{\end{equation}}

\begin{document}

\title{Scalar QED Action Density and Schwinger Pair Production in ${\rm (A)dS}_2$}
\author{Sang Pyo Kim$^{a,b,*}$}

\address{$^a$Department of Physics, Kunsan National University,\\ Kunsan 573-701, Korea\\
$^b$Center for Relativistic Laser Science, Institute for Basic Science (IBS),\\ Gwangju 500-712, Korea\\
$^*$E-mail: sangkim@kunsan.ac.kr}

\begin{abstract}
We review the one-loop effective action in scalar QED and the Schwinger effect in a uniform electric field in a two-dimensional (anti-) de Sitter space. The Schwinger effect has a thermal interpretation in terms of the effective temperature introduced by Cai and Kim. We propose a method to find the density of states for the charged scalar and obtain the QED action density and the pair-production rate in the in-out formalism.
\end{abstract}


\bodymatter


\section{Introduction}

The de Sitter (dS) and anti-de Sitter (AdS) space has a constant curvature with the maximal symmetry of a given dimension. The (A)dS has thus attracted attention in understanding a quantum nature of spacetime and been applied to different physics. The present accelerating universe with a cosmological constant is an asymptotically pure dS space and the early inflationary universe also underwent a quasi-exponential expansion. The pure dS space has a cosmological horizon and emits the dS radiation with the Gibbons-Hawking temperature.\cite{Gibbons-Hawking} The Einstein equation for the dS space can be explained by the black hole thermodynamics.\cite{Cai-Kim05} An intriguing feature of the dS radiation is the solitonic nature in the global coordinates in any odd-dimensional dS space.\cite{Polyakov} The discrimination of dimensions for dS radiation can be explained by the Stokes phenomenon, in which a pair of instanton actions interferes destructively in odd dimensions and constructively in even dimensions.\cite{Kim10,Kim13} The Stokes phenomenon can also be explained by the coherent destructive or constructive interference of the superadiabatic particle number.\cite{Dabrowski-Dunne}

The one-loop effective action in an electromagnetic field and in a curved spacetime reveals the interplay between quantum electrodynamics (QED) and quantum nature of curved spacetime. A strong electromagnetic field changes the quantum vacuum structure,\cite{Heisenberg-Euler,Schwinger} and an electric field with/without a parallel magnetic field pulls out virtual particles from the Dirac sea and creates pairs of particles and antiparticles known as the Schwinger effect.\cite{Schwinger} The Schwinger effect in (A)dS space entails the pair production by the electric field and the dS radiation\cite{Garriga,Kim-Page08,Kim14} or the suppression by the negative curvature of AdS space.\cite{Pioline-Troost,Kim-Page08} The one-loop effective action and the Schwinger effect has been comprehensively studied in Ref. \refcite{Cai-Kim14}. The $s$-wave of a scalar field in the Nariari geometry of a rotating dS black hole is equivalent to a charge in a uniform electric field in ${\rm dS}_2$.\cite{Anninos-Hartman,Anninos-Anous} Interestingly, the near-horizon geometry ${\rm AdS}_2 \times S^{d-1}$ of an extremal or near-extremal black hole deduces that the emission of charges from the black hole is related to the Schwinger effect in ${\rm AdS}_2$.\cite{Chen12,Chen15a,KLY15,Kim15a,Kim15b,Chen15b} The fermionic current due to the Schwinger effect\cite{FGKSSTV,Stahl-Strobel,SSX} and the holographic Schwinger effect\cite{FNPT} have been studied in the dS space.

\begin{figure}[h]
\begin{center}
\includegraphics[width=3.0in]{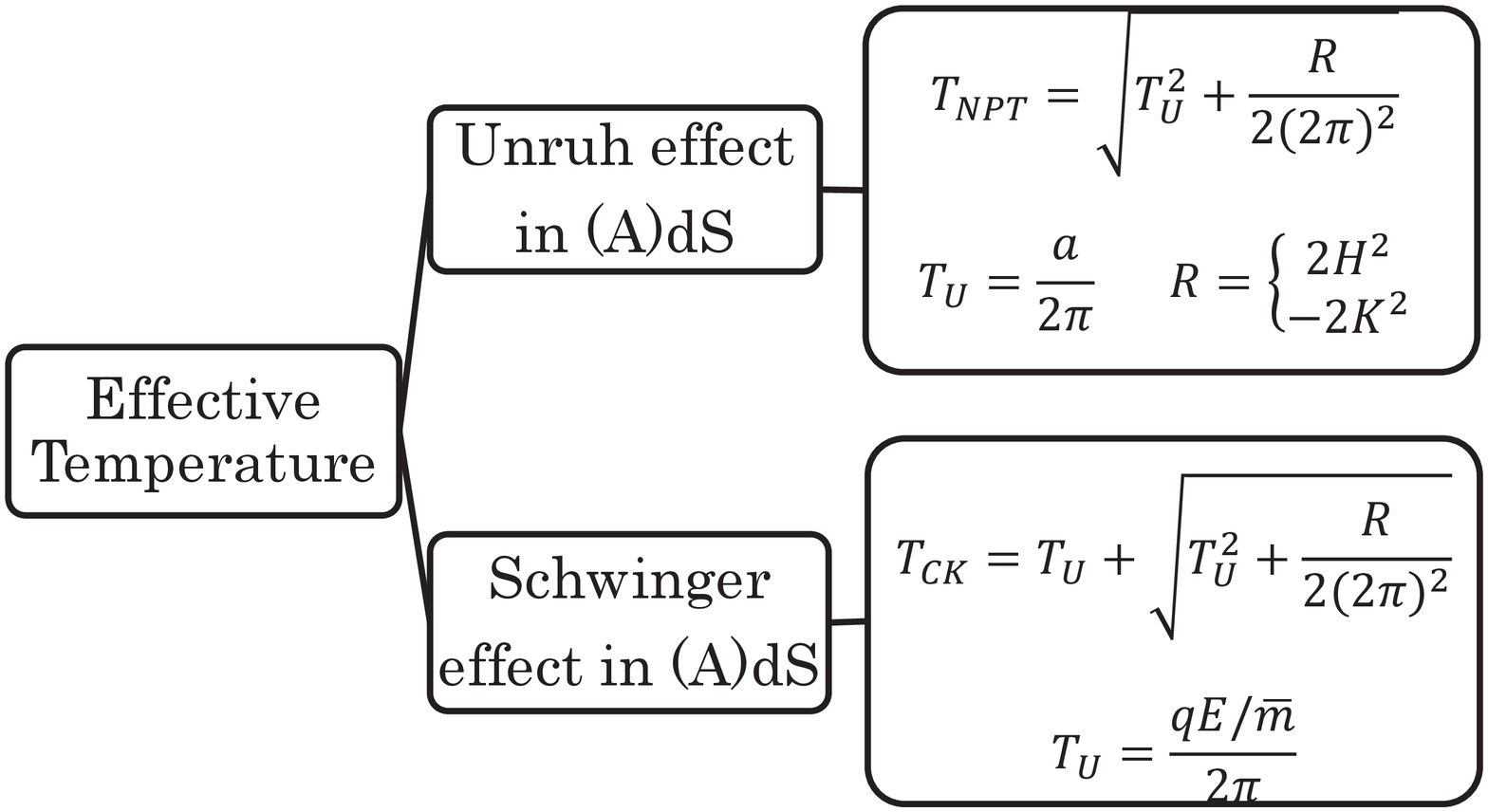}
\end{center}
\caption{The effective temperature of an accelerating observer in an (A)dS \cite{NPT,Deser-Levin}and a charged scalar in a constant electric field in an (A)dS.\cite{Cai-Kim14}}
\label{fig1}
\end{figure}

The production of particles can be explained by the (effective) temperature. The Hawking radiation has the Bose-Einstein or Fermi-Dirac distribution with the Hawking temperature and the dS radiation has a distribution, whose leading Boltzmann factor can be given by the Gibbons-Hawking temperature. The Schwinger effect in a constant or pulsed electric field has an effective temperature given by the inverse of the period of the instanton action in the Euclidean time.\cite{Hwang-Kim} Narnhofer, Peter and Thirring have shown that the effective temperature for an accelerating observer in dS space is the geometric mean of the Unruh temperature and the Gibbons-Hawking temperature,\cite{NPT} which is extended to the AdS space by Deser and Levin,\cite{Deser-Levin} as shown in Fig. \ref{fig1}. Recently, Cai and Kim have shown that the Schwinger effect in (A)dS space also has a thermal interpretation in terms of an effective temperature for the Unruh effect for accelerating charge by the electric field and the curvature effect,\cite{Cai-Kim14} as summarized in Fig. \ref{fig1}. This implies that the Schwinger effect from an extremal black hole, in which the Hawking radiation is suppressed due to the zero temperature, may also have such a thermal interpretation and similarly for a near-extremal black hole.\cite{KLY15,Kim15a,Kim15b}

\begin{figure}
\begin{center}
\includegraphics[width=2.0in]{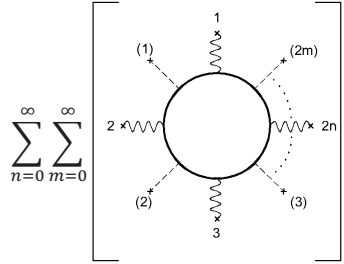}
\end{center}
\caption{One-loop Feynman diagrams: the wavy lines denote photons from the background electromagnetic field and the dashed lines denote gravitons from the curved spacetime.}
\label{fig2}
\end{figure}

In this paper we elaborate the one-loop effective action of scalar QED in Ref. \refcite{Cai-Kim14} and also advance a method to find the density of states for the one-loop action and the Schwinger effect. In particular, the density of states is important to obtain the one-loop QED action density from the integrated one-loop action in the in-out formalism and the rate of pair production from the mean number by the Bogoliubov transformation. The one-loop effective action is equivalent to all the one-loop diagrams with even number of photons and gravitons, as shown in Fig. \ref{fig2}, whose summation is beyond the computational practicality. However, in the in-out formalism by Schwinger and DeWitt, the one-loop action is derived from the scattering amplitude between the in- and out-vacua\cite{Dewitt03}
\begin{eqnarray}
e^{i {\cal W}} = e^{i \int (-g)^{1/2} d^2x {\cal L}_{\rm eff}} = \langle 0, {\rm out} \vert 0, {\rm in} \rangle. \label{in-out scat}
\end{eqnarray}
In fact, the propagator or Green function of the charge in the electric field and in the (A)dS space gives the scattering amplitude.

\section{In- and Out-Vacuum Solutions and Bogoliubov Coefficients in ${\rm (A)dS}_2$}

We study the one-loop effective action of a scalar with mass $m$ and charge $q$ in a uniform electric field in the planar coordinates of ${\rm dS}_2$ space
\begin{eqnarray}
ds^2 = -dt^2 + e^{2Ht}dx^2, \quad A_1 (t) = - (E/H) (e^{Ht} -1), \label{ds gauge}
\end{eqnarray}
and in ${\rm AdS}_2$ space
\begin{eqnarray}
ds^2 = - e^{2Kx} dt^2 + dx^2, \quad A_0 (t) = - (E/K) (e^{Kx} -1). \label{ads gauge}
\end{eqnarray}
The one-loop action differs from quantum electrodynamic interaction by summing Feynman diagrams.
The gauge potentials are chosen to have the Minkowski space limit ($H = 0$ or $K=0$) and the scalar curvature is ${\cal R} = 2 H^2 (-2K^2)$.
Then the momentum mode $\phi_{k} = e^{-Ht/2} \varphi_{k}$ in ${\rm dS}_2$ has the equation
\begin{eqnarray}
\ddot{\varphi}_{k} (t) + H^2 \Bigl(\gamma_{dS}^2 + \bar{k}^2 e^{-2Ht} + 2 \frac{qE}{H^2} \bar{k} e^{-Ht} \Bigr) \varphi_{\bf k} (t) = 0, \label{ds mod}
\end{eqnarray}
where the dimensionless parameters are
\begin{eqnarray}
\gamma_{dS}^2 =  \Bigl(\frac{qE}{H^2} \Bigr)^2 + \Bigl( \frac{m}{H} \Bigr)^2  -  \frac{1}{4}, \quad \bar{k} (k) = \frac{k}{H} - \frac{qE}{H^2}. \label{ds freq}
\end{eqnarray}
On the other hand, the energy mode $\phi_{\omega} = e^{-Kx/2} \varphi_{\omega}$ in the ${\rm AdS}_2$ space is given by the equation
\begin{eqnarray}
{\varphi}''_{\omega} (x) + K^2 \Bigl(\gamma_{AdS}^2 + \bar{\omega}^2 e^{-2Kx} - 2 \frac{qE}{K^2} \bar{\omega} e^{-Kx} \Bigr) (x) \varphi_{\omega} (x) = 0, \label{ads mod}
\end{eqnarray}
where another set of dimensionless parameters
\begin{eqnarray}
\gamma^2_{AdS} = \Bigl(\frac{qE}{K^2} \Bigr)^2 - \Bigl(\frac{m}{K} \Bigr)^2 - \frac{1}{4}, \quad \bar{\omega} (\omega) = \frac{\omega}{K} - \frac{qE}{K^2}.  \label{ads freq}
\end{eqnarray}

In the ${\rm dS}_2$ space, the positive and negative solutions for the in-vacuum ($t = - \infty$) are given by the Whittaker function as
\begin{eqnarray}
\phi^{(+)}_{{\rm in}, k} (t) &=& \frac{e^{\pi \bar{k}_0/2}}{\sqrt{2 H \bar{k}}} W_{- i \bar{k}_0, i \gamma_{dS}} (2 e^{-i \pi/2} \bar{k} e^{-Ht}), \nonumber\\
\phi^{(-)}_{{\rm in}, k} (t) &=& \frac{e^{\pi \bar{k}_0/2}}{\sqrt{2 H \bar{k}}} W_{i \bar{k}_0, i \gamma_{dS}} (2 e^{i \pi/2} \bar{k} e^{-Ht}), \label{ds in-sol}
\end{eqnarray}
with $\bar{k}_0 = \bar{k} (k = 0)$ and those for the out-vacuum ($t = \infty$) are
\begin{eqnarray}
\phi^{(+)}_{{\rm out}, k} (t) &=& \frac{e^{- i \pi/4} e^{\pi \gamma_{dS}/2}}{\sqrt{4 H \bar{k} \gamma_{dS}}} M_{i \bar{k}_0, i \gamma_{dS}} (2 e^{i \pi/2} \bar{k} e^{-Ht}), \nonumber\\
\phi^{(-)}_{{\rm out}, k} (t) &=& \frac{e^{- i \pi/4} e^{- \pi \gamma_{dS}/2}}{\sqrt{4 H \bar{k} \gamma_{dS}}} M_{i \bar{k}_0, - i \gamma_{dS}} (2 e^{i \pi/2} \bar{k} e^{-Ht}). \label{ds out-sol}
\end{eqnarray}
The solutions (\ref{ds in-sol}) and (\ref{ds out-sol}) are chosen to satisfy the quantization rule $e^{Ht} {\rm Wr}_{(t)} [\phi^{(+)}_{k} (t),\phi^{(-)}_{k} (t)] = i$. Similarly, the solutions for the in-vacuum ($x = - \infty$) for the tunneling boundary condition are
\begin{eqnarray}
\phi^{(+)}_{{\rm in}, \omega} (x) &=& \frac{e^{-i \pi/2} e^{\pi \bar{\omega}_0/2}}{\sqrt{2 K \bar{\omega}}} W_{- i \bar{\omega}_0, - i \gamma_{AdS}} (2 e^{i \pi/2} \bar{\omega} e^{-Kx}), \nonumber\\
\phi^{(-)}_{{\rm in}, \omega} (x) &=& \frac{e^{-i \pi/2} e^{- \pi \bar{\omega}_0/2}}{\sqrt{2 K \bar{\omega}}} W_{i \bar{\omega}_0, - i \gamma_{AdS}} (2 e^{-i \pi/2} \bar{\omega} e^{-Kx}), \label{ads in-sol}
\end{eqnarray}
with $\bar{\omega}_0 = \bar{\omega} (\omega = 0)$ and those for the out-vacuum ($x = \infty$) are
\begin{eqnarray}
\phi^{(+)}_{{\rm out}, \omega} (x) &=& \frac{e^{-i \pi/4} e^{- \pi \gamma_{AdS}/2}}{\sqrt{4 K \bar{\omega} \gamma_{AdS}}} M_{- i \bar{\omega}_0, - i \gamma_{AdS}} (2 e^{i \pi/2} \bar{\omega} e^{-Kx}), \nonumber\\
\phi^{(-)}_{{\rm in}, \omega} (x) &=& \frac{e^{-i \pi/4} e^{\pi \gamma_{AdS}/2}}{\sqrt{4 K \bar{\omega} \gamma_{AdS}}} M_{- i \bar{\omega}_0, i \gamma_{AdS}} (2 e^{i \pi/2} \bar{\omega} e^{-Kx}). \label{ads out-sol}
\end{eqnarray}
The solutions (\ref{ads in-sol}) and (\ref{ads out-sol}) are chosen to satisfy the quantization rule $e^{Kx} {\rm Wr}_{(x)} [\phi^{(-)}_{\omega} (x),\phi^{(+)}_{\omega} (x)] = i$ for the tunneling boundary condition.\cite{Kim-Page06}

\section{Thermal Interpretation of Schwinger Effect in ${\rm (A)dS}_2$}

Using (\ref{ds in-sol}) and (\ref{ds out-sol}), we recalculate the Bogoliubov coefficients in the ${\rm dS}_2$ space\cite{Cai-Kim14}
\begin{eqnarray}
\alpha_{k} &=& \frac{e^{- i \pi/4}}{\sqrt{2 \gamma_{dS} }} e^{\pi(\gamma_{dS} + \bar{k}_0)/2}  \frac{\Gamma(1+ 2i \gamma_{dS})}{\Gamma (\frac{1}{2} + i \gamma_{dS} - i \bar{k}_0)}, \nonumber\\
\beta_{k} &=&  \frac{e^{3i\pi/4}}{\sqrt{2 \gamma_{dS} }} e^{\pi(\bar{k}_0 - \gamma_{dS})/2}  \frac{\Gamma(1+ 2i \gamma_{dS})}{\Gamma (\frac{1}{2} + i \gamma_{dS} +i \bar{k}_0)}, \label{ds bog}
\end{eqnarray}
and from Eqs. (\ref{ads in-sol}) and (\ref{ads out-sol}), we find the Bogoliubov coefficients in ${\rm AdS}_2$ space
\begin{eqnarray}
\alpha_{\omega} &=&  e^{\pi \gamma_{AdS}}  \frac{\Gamma(\frac{1}{2}- i \gamma_{AdS} + i \bar{\omega}_0)}{\Gamma (\frac{1}{2}- i \gamma_{AdS} - i \bar{\omega}_0)}, \nonumber\\
\beta_{\omega} &=&  e^{- i \pi/4} e^{\pi (\gamma_{AdS}+ \bar{\omega}_0)/2}  \frac{\Gamma(\frac{1}{2}- i \gamma_{AdS} + i \bar{\omega}_0)}{\Gamma (\frac{1}{2}- 2 i \gamma_{AdS})}.\label{ads bog}
\end{eqnarray}
Hence, the mean numbers of pairs by the Schwinger effect are
\begin{eqnarray}
N_{\rm dS} = \vert \beta_{k} \vert^2 = \frac{e^{- 2 \pi (\gamma_{dS} + \bar{k}_0)} + e^{- 4 \pi \gamma_{dS}}}{1+ e^{-4 \pi \gamma_{dS}}}, \label{ds pair}
\end{eqnarray}
\begin{eqnarray}
N_{\rm AdS} = \vert \beta_{\omega} \vert^2 = \frac{e^{2 \pi (\gamma_{AdS} + \bar{\omega}_0)} + e^{- 2 \pi (\gamma_{AdS} -  \bar{\omega}_0 )}}{1+ e^{-2 \pi ( \gamma_{AdS} - \bar{\omega}_0)}}. \label{ads pair}
\end{eqnarray}
The binding nature of ${\rm AdS}_2$ provides an effective mass $\bar{m} = m\sqrt{1+ (K/2m)^2}$ and thus increases the critical strength $E_{\rm C} = \bar{m}^2/q$ while the separating nature of ${\rm dS}_2$ gives $\bar{m} = m\sqrt{1- (H/2m)^2}$ and thus lowers the critical strength. Note
the Breitenlohler-Freedman bound $\gamma_{\rm AdS}^2 <0$ for the stability of AdS against Schwinger emission.

Recently, Cai and Kim have proposed a thermal interpretation of the Schwinger effect in the ${\rm (A)dS}_2$ space. Note that the mean number for the dS space is determined by the actions\cite{Cai-Kim14}
\begin{eqnarray}
{\cal S}_{\rm dS} = 2 \pi \bigl(\gamma_{\rm dS} + \bar{k}_0 \bigr) = \frac{\bar{m}}{T_{\rm dS}}, \quad
\bar{\cal S}_{\rm dS} = 4 \pi \gamma_{\rm dS} = \frac{\bar{m}}{\bar{T}_{\rm dS}}.
\end{eqnarray}
Here the effective temperature and the associated temperature are
\begin{eqnarray}
T_{\rm dS} = T_{\rm U} + \sqrt{T_{\rm U}^2 + T_{\rm GH}^2}, \quad \bar{T}_{\rm dS} = \frac{T_{\rm GH}^2}{2 \sqrt{T_{\rm U}^2 + T_{\rm GH}^2}}
\end{eqnarray}
where the Gibbons-Hawking temperature and the Unruh temperature for the accelerating charge by the electric field are, respectively,
\begin{eqnarray}
T_{\rm GH} = \frac{H}{2 \pi}, \quad T_{\rm U} = \frac{qE/\bar{m}}{2 \pi}.
\end{eqnarray}
On the other hand, the mean number for the AdS space is determined by the actions\cite{Cai-Kim14}
\begin{eqnarray}
{\cal S}_{\rm AdS} &=& 2 \pi \bigl(\gamma_{\rm AdS} + \bar{\omega}_0 \bigr) = \frac{\bar{m}}{T_{\rm AdS}},\nonumber\\
\bar{\cal S}_{\rm dS} &=& 2 \pi \bigl(\gamma_{\rm AdS} - \bar{\omega}_0 \bigr) = \frac{\bar{m}}{\bar{T}_{\rm AdS}},
\end{eqnarray}
where
\begin{eqnarray}
T_{\rm AdS} = T_{\rm U} + \sqrt{T_{\rm U}^2 - \Bigl(\frac{K}{2 \pi} \Bigr)^2}, \quad \bar{T}_{\rm AdS} = T_{\rm U} - \sqrt{T_{\rm U}^2 - \Bigl(\frac{K}{2 \pi} \Bigr)^2}.
\end{eqnarray}
In the Minkowski limit $(H = K = 0)$, the mean number reduces to $N_{\rm Min} = e^{- m/ 2 T_{\rm U}}$, as expected. The factor of two is intrinsic nature of the Schwinger effect.\cite{Hwang-Kim}

\section{Density of States}

The density of states is necessary to derive the QED action density and the pair-production rate in the in-out formalism. The energy-momentum integral of the off-shell mean number gives the density of states, $qE/(2 \pi)$, for a uniform electric field in the Minkowski spacetime.\cite{KLY11} In the case of a uniform magnetic field, the wave packet of Landau states gives the density of states, $qB/(2 \pi)$.

We now propose another method to find the density of states in the in-out formalism. For that purpose, we note that the field has the Fourier integral
\begin{eqnarray}
\phi^{(+)}_{\rm in} (t,x) = \int \frac{dk}{2 \pi} \phi^{(+)}_{{\rm in}, k} (t). \label{fourier}
\end{eqnarray}
Using the asymptotic form $M_{\mu, \nu} (z) = e^{-z/2} z^{1/2 + \nu}$ and changing of variable $dk = H d \bar{k}$, we rewrite Eq. (\ref{fourier}) as
\begin{eqnarray}
\phi^{(+)}_{\rm in} (t,x) = H \int \frac{d \bar{k}}{2 \pi} e^{\theta (\bar{k}) (t)} f^{(+)}_{\rm in} (t),
\end{eqnarray}
where $f^{(+)}_{\rm in} (t)$ is a function independent of $\bar{k}$, and
\begin{eqnarray}
\theta (\bar{k}) (t) = - i \bar{k} e^{-Ht} + i \gamma_{dS} \ln (\bar{k}).
\end{eqnarray}
The Gaussian integral from the saddle point method gives
\begin{eqnarray}
\phi^{(+)}_{\rm in} (t,x) = \frac{H \sqrt{\gamma_{dS}}}{\sqrt{2 \pi}} f^{(+)}_{\rm in} (t). \label{saddle}
\end{eqnarray}
In the tunneling picture, the decay of the vacuum due to the Schwinger effect is determined by the magnitude square of the field
\begin{eqnarray}
\vert \phi^{(+)}_{\rm in} (t,x) \vert^2 = \frac{H^2 \gamma_{dS}}{2 \pi} \vert f^{(+)}_{\rm in} (t) \vert^2,
\end{eqnarray}
and the momentum independent part $\vert f^{(+)}_{\rm in} (t) \vert^2$ determines the mean number of created pairs.

Therefore, the density of states is given by
\begin{eqnarray}
{\cal D}_{\rm dS} = \frac{H^2 \gamma_{\rm dS}}{2 \pi}, \quad {\cal D}_{\rm AdS} = \frac{K^2 \gamma_{\rm AdS}}{2 \pi}. \label{den st}
\end{eqnarray}
In the Minkowski spacetime $(H=K=0)$ the density of states becomes ${\cal D} = qE/(2 \pi)$. Finally, we obtain the pair-production rate per unit two-volume
\begin{eqnarray}
\frac{d N_{\rm dS}}{dt dx} = \frac{H^2 \gamma_{\rm dS}}{2 \pi} \vert \beta_{k} \vert^2, \quad \frac{d N_{\rm AdS}}{dt dx} = \frac{K^2 \gamma_{\rm AdS}}{2 \pi} \vert \beta_{\omega} \vert^2.
\end{eqnarray}
The density of state has been proposed for spinor QED in the dS space.\cite{Haouat-Chekireb}

\section{QED Action Density in ${\rm (A)dS}_2$}

From the integrated one-loop action (\ref{in-out scat}) follows the QED action density, which is given by the Bogoliubov coefficient, which is in turn expressed by gamma functions with complex arguments
\begin{eqnarray}
{\cal L} = i {\cal D} \ln (\alpha^* ) = i {\cal D} \sum_{l} \ln \Gamma (a_{l} - i b_{l}).
\end{eqnarray}
To obtain the QED action density, one uses the gamma function in the proper-time integral
\begin{eqnarray}
\ln (\Gamma (a_l - i b_l)) = \int^{\infty}_{0} \frac{ds}{s} \Bigl[ \frac{e^{-(a_l - ib_l) s}}{1- e^{-s}} - \cdots \Bigr], \label{gamma fn}
\end{eqnarray}
and then applies the Schwinger subtraction scheme to regulate divergent terms via the renormalization of the vacuum energy and charge.\cite{KLY}
Remarkably, the consistence relation, $2 {\rm Im} {\cal W} = \ln ( 1 + \vert \beta \vert^2)$, between the imaginary part of the effective action and the pair-production rate, is the consequence of the Cauchy's residue theorem applied to the gamma function (\ref{gamma fn}). Finally, we find the QED action density in the ${\rm (A)dS_2}$ space
\begin{eqnarray}
{\cal L}^{(1)}_{\rm dS} &=& \frac{H^2 \gamma_{\rm dS}}{2(2 \pi)} \int_{0}^{\infty} \frac{ds}{s} \Bigl[ e^{ -(\gamma_{\rm dS} + \bar{k}_0)s} \Bigl(\frac{1}{\sin(s/2)} - \frac{2}{s} - \frac{s}{12} \Bigr) \nonumber\\&& - e^{ - 2 \gamma_{\rm dS}s} \Bigl(\frac{\cos(s/2)}{\sin(s/2)} - \frac{2}{s} +\frac{s}{6} \Bigr) \Bigr], \nonumber\\
{\cal L}^{(1)}_{\rm AdS} &=& \frac{K^2 \gamma_{\rm AdS}}{2(2 \pi)} \int_{0}^{\infty} \frac{ds}{s} e^{ -\gamma_{\rm AdS}s} \cosh(\bar{\omega}_0 s)
\Bigl(\frac{1}{\sin(s/2)} - \frac{2}{s} - \frac{s}{12} \Bigr). \label{(a)ds qed}
\end{eqnarray}

\section{Conclusion}

The scalar QED action and Schwinger effect has been recalculated in the ${\rm (A)dS_2}$ space. A new method has been advanced to find the density of states and to obtain the QED action density and the Schwinger pair-production rate in the ${\rm (A)dS_2}$ space, which is the main result.

\section*{Acknowledgments}
The author thanks Christian Schubert for useful discussion on the worldline instantons in de Sitter space and Won Kim for drawing the figure.
This work was supported by IBS (Institute for Basic Science) under IBS-R012-D1.

\end{document}